\documentclass[11pt,onecolumn]{article}

\usepackage[utf8]{inputenc}
\usepackage[T1]{fontenc}
\usepackage{lmodern}

\usepackage[american]{babel}       
\usepackage{microtype}
\microtypesetup{tracking=true}
\usepackage[margin=1in]{geometry}
\usepackage{setspace}
\usepackage{parskip}

\usepackage{amsmath}
\numberwithin{equation}{section}
\usepackage{amssymb}
\usepackage{amsfonts}
\usepackage{mathtools}
\usepackage{bm}
\usepackage{siunitx}

\usepackage{graphicx}
\usepackage{booktabs}
\usepackage{multirow}
\usepackage{array}

\usepackage{xcolor}
\usepackage{listings}
\lstdefinestyle{FinanceCode}{
  language=R,
  basicstyle=\ttfamily\small,
  commentstyle=\color{gray},
  keywordstyle=\color{blue},
  stringstyle=\color{green!50!black},
  numberstyle=\tiny\color{gray},
  breakatwhitespace=false,
  breaklines=true,
  captionpos=b,
  keepspaces=true,
  numbers=left,
  numbersep=5pt,
  showspaces=false,
  showstringspaces=false,
  showtabs=false,
  tabsize=2
}
\usepackage{url}
\renewcommand{\UrlFont}{\normalfont}

\usepackage{footmisc}

\usepackage[style=apa,backend=biber,natbib=true]{biblatex} 
\DeclareLanguageMapping{american}{american-apa}
\usepackage[colorlinks=true,
            linkcolor=blue,
            citecolor=blue,
            urlcolor=blue]{hyperref}

\lstdefinestyle{FinanceCode}{
    language=R,
    basicstyle=\ttfamily\small,
    commentstyle=\color{gray},
    keywordstyle=\color{blue},
    stringstyle=\color{green!50!black},
    numberstyle=\tiny\color{gray},
    breakatwhitespace=false,
    breaklines=true,
    captionpos=b,
    keepspaces=true,
    numbers=left,
    numbersep=5pt,
    showspaces=false,
    showstringspaces=false,
    showtabs=false,
    tabsize=2
}
\title{\textbf{A New Application of Hoeffding's Inequality \\ 
Can Give Traders Early Warning of Financial Regime Change}}

\author{
\centering
by Daniel Egger\textsuperscript{1} \; and \; Jacob Vestal\textsuperscript{2}
}

\date{}

\begin{document}

\maketitle

\begingroup
\renewcommand\thefootnote{\arabic{footnote}}
\footnotetext[1]{Pratt School of Engineering, Duke University. Email: \texttt{daniel.egger@duke.edu}}
\footnotetext[2]{Pratt School of Engineering, Duke University. Email: \texttt{jacob.vestal@duke.edu}}
\endgroup

\vspace{5mm}
% \noindent \textbf{Keywords:} Financial Modeling, Asset Pricing, [Keyword 1], [Keyword 2] \\
% \textbf{JEL Classification:} G11, G12, C58

\newpage

\section{Introduction}
\label{sec:intro}
Hoeffding's Inequality defines the maximum probability that an observed mean $\bm{\bar{X}}$ on $\bm{N}$ samples of an independent random variable (RV) varies from the RV’s true long-term mean $\bm{\mu}$ by at least a specified tolerance $\bm{t}$. It applies to all RVs that are bounded; that is to say, RVs for which any measurement $X_i$ of the RV is confined to the interval $\bm{a \leq X_i \leq b, a, b \in \mathbb{R}.}$ For the specific case that $\bm{a = 0, b = 1}$:

\begin{equation}
\label{eq:hoeffding}
\mathbb{P}[\bar{X} - \mu \geq t] \leq \boldsymbol{e^{-2t^2N}}
\refstepcounter{equation}\tag*{\textbf{(\theequation)}}
\end{equation}

\noindent Hoeffding’s Inequality holds true regardless of the underlying distribution of the RV. It holds whether successive observations of the RV are independent or serially correlated. Hoeffding’s Inequality serves as a useful “worst-case” metric in applications where the modeler avoids assumptions about variance, autocorrelation, or the underlying distribution, as the bounds in $\bm{1.1}$ become tighter only when such assumptions are included \citep{Azuma1967, Talagrand1995, ZhangChen2021}.

\noindent In statistical learning theory, the true value of a predictive model's expected future performance  $\bm{\mu}$ on out-of-sample data cannot be known, but out-of-sample data are assumed to be drawn from the same distribution as test data, for which average performance is observed to be $\bm{\bar{X}}$, so the assumption is made that $\mathbb{E}(\bar{X}) = \bm{\mu}$.  In this context, Hoeffding’s Inequality demonstrates why learning is feasible: it constrains the risk that $\bm{\mu}$ will exceed $\bm{\bar{X}}$ by at least tolerance $\bm{t}$ by defining the worst-case, maximum value for the probability $\mathbb{P}[|\bar{X} - \bm{\mu}| \geq \bm{t}]$. This probability maximum falls exponentially as $\bm{N}$ and $\bm{t^2}$ increase.

\noindent Here we propose an alternative, \textit{trader's application} of Hoeffding's Inequality. 
\par
We define a \textit{financial regime} as the complex of largely unknown causal relationships between the distributions of obtainable signals and future asset prices that support repeated successful application of a given trading strategy. The often-unspoken underlying hypothesis, $\bm{H_0}$, needed to justify continued application of a given strategy is that the same financial regime under which the strategy was found to work remains available.  
\par
 \noindent Traditionally, financial regimes have been thought of in terms of structural shifts in basic characteristic variables such as volatility \citep{Hamilton1989}, stock returns \citep{SchallerVanNorden1997}, or interest rates \citep{AngBekaert2002}. This formulation has enabled the development of forward-looking regime-switching models that attempt to forecast changes in regimes in macroeconomic indices such as the S\&P 500 and the VIX \citep{Xu2024}.
\par
\noindent Our definition of “regime change” departs from the traditional view in that it places a practical emphasis on the efficacy of a trading strategy as realized by the trader. \textit{Financial regime change} is defined here to mean that the underlying distributions and causal patterns have changed enough to render a trading strategy ineffective. This is the alternative hypothesis $\bm{H_1}$.
\par
\noindent The trader's application uses Hoeffding’s Inequality as a regime change indicator; in the trader's application, $\bm{\mu_{i,j}}$ a trader's expected future performance of a trading strategy during a specified forward-looking time interval $\bm{[i,j]}$, conditional on the hypothesis $\bm{H_0}$ that the favorable financial regime persists. 
\par
Applying Hoeffding's Inequality requires only that a trader commit to a specific belief, $\bm{\mu_{i,j}}$ that quantifies performance sufficient to offset the opportunity cost or risk of a given strategy. The belief can be expressed either in negative terms, as a loss percentage, or in positive terms as a win percentage or percent return on equity, so long as all trades within the strategy are structured such that their performance is bounded. Committing to a belief allows the trader to make inferences from the realized difference $\bm{t_{i,j} = \bar{X}_{i,j} - \mu_{i,j}}$ between a trading strategy’s actual measured performance and its expected performance during time interval $\bm{[i,j]}$. If $\bm{t_{i,j}}$ increases, falling maximum values for $\bm{{\mathbb{P}[|\bar{X} - \mu | \geq \varepsilon}]}$ indicate decreasing plausibility of $\bm{H_0}$ and increasing plausibility of the alternative hypothesis of regime change, $\bm{H_1}$.

\textbf{Statistical Learning Theory Application of Hoeffding's Inequality}

The widely known application of Hoeffding’s Inequality is motivated by the need to quantify whether a given machine learning model has been trained and tuned sufficiently for its output to be trusted when applied to new data in a production environment. Hoeffding's Inequality is adopted to quantify the mathematical relationship between the empirical error rate of a machine learning classifier $\bm{E_{in}}$, and that classifier's unknown future error probability $\bm{E_{out}}$:

\begin{equation}
\label{eq:hoeffdingReplacing}
\bm{
\mathbb{P}[|E_{in} - E_{out}| > \varepsilon] \leq 2e^{-2\varepsilon^2N}
}
\tag*{\textbf{(1.2)}}
\end{equation}

\begin{center}
\textit{Hoeffding’s Inequality, replacing $\ge$ with $>$ and tolerance $t$ with estimation error $\bm{\varepsilon}$.}\\
See \citep[p.~22]{AbuMostafa2012}.

\end{center}

This application is based on the insight that the error rate of a classifier is itself a bounded random variable described by \textbf{Inequality 1.1}; by definition, an error rate cannot fall below \textbf{0} or exceed \textbf{1}.
\par
\textbf{Inequality 1.2} helps to quantify the risks of shipping a new predictive model by providing an upper bound on the probability that the difference between the empirical error rate $\bm{E_{in}}$ on a set of $\bm{N}$ observed examples and $\bm{E_{out}}$ the true error probability, exceeds a tolerance $\bm{\varepsilon}$. 
\par
Limiting the inequality to those cases where $\bm{E_{out} > E_{in}}$ (because in practice the concern is primarily that a classifier may in future performs worse than it did on a test sample) reduces the probability of exceeding the tolerance by half, so \textbf{Inequality 1.2} can be restated: 

{\boldmath
\begin{equation}
\label{eq:hoeffdingRestated}
\mathbb{P}[|E_{in} - E_{out}| > \varepsilon] \leq e^{-2\varepsilon^2N} 
\tag{\textbf{1.3}}
\end{equation}
\unboldmath
}

Note the difference in kind between $\bm{E_{in}}$ and $\bm{E_{out}}$. $\bm{E_{in}}$ is an observed mean, $\bm{\bar{X}}$, of bounded random variables based on a sample of $N$ independent draws, while $\bm{E_{out}}$  is an expectation $\bm{\mu}$ of an unknown stationary distribution from which all draws $\bm{X_i}$ must be made. Although $\bm{\mu}$ is the cause of $\bm{\bar{X}}$, in this learning theory application of the Inequality it is only through measurement of $\bm{\bar{X}}$  that one infers an estimate for unknown $\bm{\mu}$.
\par
In the statistical learning theory context, Hoeffding's Inequality is always applied under the assumption that the $\bm{E_{in}}$ = $\bm{\bar{X}}$ is calculated from N observations drawn from an unknown stationary distribution with expectation $\bm{E_{out}}$ = $\bm{\mu}$ . The true value of $\bm{\mu}$ is then inferred as approximately equal to the in-sample mean $\bm{\bar{X}}$, and combining the assumption with the inference, a maximum probability of $\bm{\mu}$ meeting or exceeding any given tolerance $\bm{\varepsilon}$ can be calculated. 
\par
However, nothing in the original formulation of Hoeffding's Inequality requires this relationship between $\bm{\bar{X}}$ and $\bm{\mu}$. 
\par

% would you prefer for this to be a new page? 
\newpage

\textbf{Equivalent forms of Hoeffding’s Inequality}

Although AML correctly (more conservatively) give the probability bound in cases where the difference is greater than the chosen tolerance, Hoeffding's original formulation also includes the case where the difference equals the tolerance: 

\begin{equation}
\label{eq:hoeffdingEquals}
\mathbb{P}[\bar{X} - \mu \geq t] \leq \bm{e^{-2t^2N}} 
\refstepcounter{equation}\tag*{\textbf{(1.4)}}
\end{equation}

The bounds that Hoeffding derived also apply to the probabilities of events other than ${\bar{X} - \mu \geq t}$. For example, Hoeffding’s original proof of his Inequality starts by defining a Bernoulli bounded random variable as follows: 

\begin{equation*}
\bm{V(S) = }
    \begin{cases} 
      \bm{1}               & \bm{\text{if } S - \mathbb{E}(nt) -nt \geq 0}\\ 
      \bm{0}               & \textbf{\text{otherwise}}
    \end{cases}
\end{equation*}

If instead, one begins the proof by flipping the “$\geq$” to define a new random variable:

\begin{equation*}
\bm{V(S) = }
    \begin{cases} 
      \bm{1}               & \bm{\text{if } S - \mathbb{E}(nt) -nt \leq 0}\\ 
      \bm{0}               & \textbf{\text{otherwise}}
    \end{cases}
\end{equation*}

and proceeds through the mechanics of Hoeffding’s original proof with the new definition, then the following relation is derived:

\begin{equation}
\label{eq:hoeffdingOrigWithNewDef}
\mathbb{P}[\bar{X} - \mu \leq -t] \leq \bm{e^{-2t^2N}} 
\refstepcounter{equation}\tag*{\textbf{(1.5)}}
\end{equation}

Because $\bar{X} - \mu \geq t \Longrightarrow \mu - \bar{X} < -t$  and $\bar{X}-\mu\le-t\Longrightarrow\mu-\bar{X} > t$ it follows that

\begin{equation}
\label{eq:1.6}
\mathbb{P}[\mu - \bar{X} < -t] \leq \bm{e^{-2t^2N}} 
\refstepcounter{equation}\tag*{\textbf{(1.6)}}
\end{equation}
\par
and 
\begin{equation}
\label{eq:1.7}
\mathbb{P}[\mu - \bar{X} > -t] \leq \bm{e^{-2t^2N}} 
\refstepcounter{equation}\tag*{\textbf{(1.7)}}
\end{equation}
\par

\textbf{Trading Application of Hoeffding's Inequality }
\par
Most traders will choose to base their expected performance $\bm{\mu_{i,j}}$ on back-tested historical data, perhaps in tandem with a reasonable forecasting model if any trends are expected. However, because $\bm{\mu_{i,j}}$ represents what the trader expects for the strategy’s performance, there is no formal limitation on \textit{how} a trader calculates $\bm{\mu_{i,j}}$: only that the trader \textit{commits to} $\bm{\mu_{i,j}}$ as the strategy’s expected performance under the hypothesis $\bm{H_0}$ of the market regime assumed extant in time interval \textbf{[\textit{i,j}]}.  
\par
In the trading application, instead of inferring the value of $\bm{\mu}$ from $\bm{\bar{X}}$ as in statistical learning theory, one can start with any assumption about the value of $\bm{\mu}$ from any source whatsoever, and continually observe the value of $\bm{t}$ as $\bm{t_{i,j}} = \bm{\mu_{i,j}} - \bm{\bar{X}_{i,j}}$.  Here, $\bm{\bar{X}_{i,j}}$ is the observed average performance of a series of N round-trip trades and $\bm{t_{i,j}}$ is the continually-measured difference between $\bm{\mu_{i,j}}$ and $\bm{\bar{X}_{i,j}}$. 

\newpage
So long as $\bm{\bar{X}_{i,j}}$  is drawn from the bounded stationary distribution with expectation $\bm{\mu_{i,j}}$, the relations (1.3) – (1.7) apply.  In other words, Hoeffding's Inequality can be restated as a maximum \textit{conditional probability}:

\begin{equation}
\label{eq:1.7-called1.8}
\mathbb{P}[\bar{X} - \mu \geq t \; | \; E(\bar{X}) = \mu] \leq \bm{e^{-2t^2N}} 
\refstepcounter{equation}\tag*{\textbf{(1.8)}}
\end{equation}
or 
\begin{equation}
\label{eq:1.8-called1.9}
\mathbb{P}[\bar{X} - \mu \geq t \; | \; H_0] \leq \bm{e^{-2t^2N}} 
\refstepcounter{equation}\tag*{\textbf{(1.9)}}
\end{equation}

However, since in this application $\bm{\mu}$ was not inferred from $\bm{\bar{X}}$, a falling probability indicates decreasing plausibility of the financial regime hypothesis $\bm{H_0}$, namely that the trader's underlying assumptions governing the value of $\bm{\mu}$ are correct, and increasing plausibility of the alternative hypothesis of regime change, $\bm{H_1}$. Note that this is not a claim that $\bm{e^{-2t^2N}}$ is the maximum probability the trader's assumption of $\bm{H_0}$,\ namely that E(X) = $\bm{\mu}$,\ is correct.   

% put in the detail at the footer later

\par

\textbf{Setting up a Prospective Trading Strategy with Hoeffding Signals. }
\par
\noindent\textbf{\textls[-30]{Part 1: Performance Metrics of Bounded Random Variables With Ranges Bounded by (0,1)}}\par

Any metric about the expected trade performance that is a rate between 0 and 1 can be treated as the random variable.  For example, a back-tested strategy consists of a set of trades with known entry and exit prices. These can be characterized by the following rates:
\par
$\bm{W -}$ Win percentage: what portion of trades achieved a notional gross profit, ignoring time value of money, transaction costs, etc. 
\par
$\bm{P -}$ Profit percentage: what portion of trades achieved a net profit, taking into account time value of money, transaction costs, and any other etc. 
\par
$\bm{U -}$ Target upside exit percentage: assuming the trading rules for the strategy are constructed so as to close out the position at a target percentage gain, what percentage of trades reached the target exit price? (In the absence of an automatic close-out target, the percent of all exits above a threshold upside return on equity can be tracked). 
\par
$\bm{D -}$  Stop loss percentage: assuming the trading rules for the strategy are constructed so as to close out a position automatically at a maximum percentage loss, what percentage of trades reached the stop loss price? (In the absence of an automatic stop-loss target, the percent of all exits worse than a threshold equity loss can be tracked).
\par
$\bm{M }$ (for Misc.) \textbf{-} Any rate over the range (0,1) and derived from the back-tested trade data can be tracked. 
\par
The trader defines the trading strategy in terms of entry rules, exit rules, and expected values for the above rates. After each round-trip trade is complete, the Hoeffding probabilities can be recalculated.   In one application, traders might use 50\%, 25\% and 10\% as indicators. These probabilities are the \textit{maximum probability that the difference between the actual and the target metric is due to chance alone. }
\par
When a maximum probability \textit{drops below 50\%, it is more likely than not} that the trader's current beliefs about the market regime are not completely correct. 
If a maximum probability drops below 25\%, there is a \textit{significant risk} that the trader's beliefs about the current market regime are, when taken as a whole, no longer justified.  
\par
And if a maximum probability \textit{drops below 10\%}, it is \textit{almost certain that market regime change} has occurred and it is time to halt the current trading program and rethink the trader's entire strategy in light of fundamentally changed market conditions. 
\par
Of course, traders are also free to use values other than 50\%, 25\%, and 10\%, and to automate their response to Hoeffding probability signals, as well as to use them to inform margin and leverage levels and overall stake management. 

\noindent\textbf{\textls[-30]{Part 2: Performance Metrics of Bounded Random Variables With Ranges Other Than (0,1)}}\par

\par
Hoeffding's Inequality does not require that random variables be bounded between 0 and 1, (although this is the form in which Hoeffding is invariably encountered in machine learning contexts). Where the RV is bounded by $\bm{a}$,$\bm{b}\in\mathbb{R}$ so that $\bm{a \leq X_i \leq b}$, and $\bm{b - a \neq 1}$, the formula is:

\begin{equation}
\label{eq:1.9called1.10}
\mathbb{P}[\bar{X} - \mu \geq t] \leq \bm{e^{\left(\frac{-2t^2N}{b-a}\right)}}
\refstepcounter{equation}\tag*{\textbf{(1.10)}}
\end{equation}

A significant disadvantage of using bounds broader than one is that the Hoeffding equation offers weaker signals. If traders want to use a Hoeffding signal on their mean expected continuously compounded absolute return, or mean expected continuously compounded annualized return, it is recommended that they set bounds for automatic upside exits U and stop loss downside exits D that are as close together as is consistent with their overall trading strategy.

\clearpage       
\begin{center}
  \bfseries
 BOX A
\end{center}

\textit{A typical machine learning application of Hoeffding's Inequality. The in-sample error rate in this context requires that the predictive model was not chosen or optimized with knowledge of the sample data. In other words, in this context the in-sample data is "test" data not "training" data. Hoeffding's Inequality has practical usefulness because it can then be used to obtain upper bounds for the probability of future error rates worse than the observed error rate.}

Assume an observed error rate $\bm{E_{in}}$= 5\%  in a data set of N = 1000 examples. 

By \textbf{1.6: }

\begin{equation*}
\mathbb{P}[\mu - \bar{X} > t] \leq \bm{e^{-2t^2N}} 
\end{equation*}

Therefore: 
\begin{align*}
& \mathbb{P}[\bm{(E_{out}} > 8\%] \leq 16.5\% & (\bm{\varepsilon} = 3\%, \text{ and } \mathbb{P}[\bm{(E_{out} - (E_{in} > \varepsilon] \leq e^{-2(.03)^21000}})\\
& \mathbb{P}[\bm{(E_{out}} > 9\%] \leq 4.7\% & (\bm{\varepsilon} = 4\%, \text{ and } \mathbb{P}[\bm{(E_{out} - (E_{in} > \varepsilon] \leq e^{-2(.04)^21000}}) & \text{, and}\\
& \mathbb{P}[\bm{(E_{out}} > 10\%] \leq 0.6\% & (\bm{\varepsilon} = 5\%, \text{ and } \mathbb{P}[\bm{(E_{out} - (E_{in} > \varepsilon] \leq e^{-2(.05)^21000}})\\
\end{align*}

While for the same observed error rate $\bm{E_{in}}$= 5\%  in a data set of $\bm{N}$ = 2000 examples: 

\begin{align*}
& \mathbb{P}[\bm{(E_{out}} > 8\%] \leq 2.7\% & (\bm{\varepsilon} = 3\%, \text{ and } \mathbb{P}[\bm{(E_{out} - (E_{in} > \varepsilon] \leq e^{-2(.03)^22000}})\\
& \mathbb{P}[\bm{(E_{out}} > 9\%] \leq 0.17\% & (\bm{\varepsilon} = 4\%, \text{ and } \mathbb{P}[\bm{(E_{out} - (E_{in} > \varepsilon] \leq e^{-2(.04)^22000}}) & \text{, and}\\
& \mathbb{P}[\bm{(E_{out}} > 10\%] \leq 0.0045\% & (\bm{\varepsilon} = 5\%, \text{ and } \mathbb{P}[\bm{(E_{out} - (E_{in} > \varepsilon] \leq e^{-2(.05)^22000}})\\
\end{align*}

\newpage

\begin{center}
\textbf{BOX B}
\end{center}
\begin{center}
\textit{Example of using Hoeffding's Inequality as a general regime-change indicator.}
\end{center}

\textit{A trader has performed back-testing and analysis and has arrived at the belief that a particular trading strategy’s expected success rate should be about \textbf{60\%}. However, after performing \textbf{12 }trades, the trader sees that only \textbf{42\%} were successful, a difference of 18\%. The trader uses Hoeffding’s Inequality to calculate the maximum possible probability of this difference happening by chance.}

% The trader has measured $\bar{X}- \mu \leq i -t = 0.18$ and wishes to calculate the upper bound on the probability of this happening by chance.

By \textbf{1.4}:

\begin{align*}
& \mathbb{P}[\bar{X} - \mu \geq t] \leq \bm{e^{-2t^2N}} \\
\end{align*}

By substituting .18 for 2 and 12 for N:

\begin{align*}
 \Rightarrow & \mathbb{P}[\bar{X} - \mu \geq t] \leq \bm{e^{-2 (.18)^2 12}} \\
 \Rightarrow &  \mathbb{P}[\bar{X} - \mu \geq t] \leq \bm{38.29\%} \\
\end{align*}

If the expected success rate of the trader’s strategy is correctly 60\%, then the maximum possible probability of observing a success rate of 42\% in 12 trades is about 38.29\%... which is less than 50\%. In other words, it \textit{is more likely than not} that the low return probably was \textit{not} observed by random chance. Since the performance belief was based on the persistence of a specific market regime, the trader concludes that it is more likely than not that a regime change has taken place 
\par
A Trader might then choose to reduce their stake or halt the strategy and re-evaluate their market assumptions. 

\clearpage    
\begin{center}
  \bfseries
 APPENDIX A:
\end{center}

\begin{center}
\textbf{Discussion of Definitions, Assumptions, and}
\end{center}
\begin{center}
\textbf{an Edited Proof of Hoeffding’s Original Inequalities}
\end{center}
\par
\par
\textbf{Summary of Assumptions and Definitions:}

$X=\left[X_1,X_2,\cdots X_n\right]$ symbolizes a series of observations of $n$ random variables. The mean and variance of each random variable may differ; all that is required is that $X_i$ – the value of the $i^{th}$ observation of the RV – is constrained to the interval defined by upper and lower bounds \textit{b} and \textit{a}; in other words, $a\leq X_i\leq b$ in which \textit{b}, $a \in\mathbb{R}$ and $b>a$.
\par
If a variable is random, then there is more than one possible observation for the value of that variable. If a random variable is bounded, then all observed values $X_i$ for the RV lie within [\textit{a,b}]. Therefore, the expected value of an RV thus defined exists and lies in [\textit{a,b}].
\par
The assumption of boundedness also places limitations on the possible values for variance of the RV; namely, a max value for the variance such that $\sigma^2\leq (\mu-a)(b-\mu)$; or, for the special case that ${a}=0, {b}=1:$ $\sigma^2\leq\mu(1-\mu)$.
\par
No distribution is required or assumed for any of the random variables. The Hoeffding bounds hold regardless of distribution.
\par

\textbf{Proof of Hoeffding’s Inequality 1}

\par
\textbf{Observation:} The random variables are assumed to be independent as a worse-case condition to derive maximal bounds that contain all possible cases.

\par
We now put forth an edited version of Hoeffding’s proof of Hoeffding’s First Inequality, which was given the name Theorem 2.1 in his original paper \citep{Hoeffding1963}.

\par
From the foundations established above, we may construct the definition of a new random variable $S=\sum_{i=n}^{n}X_i$; defined as the sum of \textit{n }draws (or observations) of \textit{n} random variables. \textit{S} is also bounded, with finite first and second moments, because it is a linear combination of observations of random variables that have those properties. 

\textbf{Setup:} a Bernoulli random variable whose expectation value equals the probability of the event on which we wish to place a probability bound.
\par
Define a Bernoulli random variable V(S), as below, that returns \textbf{1} or \textbf{0} as below:
\begin{equation*}
\bm{V(S) = }
    \begin{cases} 
      \bm{1}               & \bm{\text{if } S - \mathbb{E}(S) -nt \geq 0}\\ 
      \bm{0}               & \textbf{\text{otherwise}}
    \end{cases}
\end{equation*}
\begin{align*}
& \textbf{Wherein: } \\
& \textit{\textbf{n:}} \text{ number of observations made on RV} \\
& \textit{\textbf{S:}} \text{ sum of \textit{n} observations of RV} \\
& \mathbb{E}(\bm{S}): \text{ Expected value of the sum \textit{S}} \\
& \bm{t:} \text{ tolerance, set or observed by the analyst.} \\
\end{align*}

Under this definition, no matter what values $\bm{S}$, $\mathbb{E}(\textbf{S})$, and \textit{\textbf{nt}} may take, there are exactly three cases for $\bm{S} - \mathbb{E}(S) - \bm{nt}$:

\begin{align*}
& \textbf{Case 1: } S - \mathbb{E}(S) -nt < 0 \rightarrow V(S) = \bm{0} \\
& \textbf{Case 2: } S - \mathbb{E}(S) -nt = 0 \rightarrow V(S) = \bm{1} \\
& \textbf{Case 3: } S - \mathbb{E}(S) -nt > 0 \rightarrow V(S) = \bm{1} \\
\end{align*}

Now let $\bm{h}$ be any real-valued, positive constant. Similar to the above, there are exactly three cases for the expression $\bm{e}^{\bm{h}(\bm{S} - \mathbb{E}(\bm{S}) -\bm{nt})}$:

\begin{align*}
& \textbf{Case 1: } S - \mathbb{E}(S) -nt < 0 \rightarrow e^{h(S - \mathbb{E}(S) -nt)} > \bm{0} \\
& \textbf{Case 2: } S - \mathbb{E}(S) -nt = 0 \rightarrow e^{h(S - \mathbb{E}(S) -nt)} = \bm{1} \\
& \textbf{Case 3: } S - \mathbb{E}(S) -nt > 0 \rightarrow e^{h(S - \mathbb{E}(S) -nt)} > \bm{1} \\
\end{align*}

Clearly, in all cases, for all values of $\bm{S}$, $\mathbb{E}(S)$, and \textit{\textbf{nt}}, $\bm{V(S)}$ is bounded by $\bm{e}^{\bm{h}(\bm{S} - \mathbb{E}(\bm{S}) -\bm{nt})}$ as $\bm{V(S)} \leq \bm{e}^{\bm{h}(\bm{S} - \mathbb{E}(\bm{S}) -\bm{nt})}$. Therefore, its expectation value is also bounded:

\begin{equation*}
 \mathbb{E}(\bm{V(S)}) \leq \mathbb{E}(\bm{e}^{\bm{h}(\bm{S} - \mathbb{E}(\bm{S}) -\bm{nt})})
\end{equation*}

Because $\bm{V(S)}$ equals \textbf{1} only when an observed sum exceeds its expected value by at least $\bm{nt}$ (and \textbf{0} otherwise), the expected value of $\bm{V(S)}$ equals the probability of observing a sum $\bm{S}$ that exceeds the expected sum; i.e.:

\begin{equation*}
 \mathbb{E}(\bm{V(S)}) =\mathbb{P}(\bm{S} - \mathbb{E}(\bm{S}) \geq \bm{nt})
\end{equation*}

Let $\bar{X}\equiv \bm{S/n}$ be the observed (sample) average of the sum $S$ of $n$ measurements of the RV.
\par
Let $ \mathbb{E}(\bm{\bar{X}}) = \mathbb{E}\bm{(S)/n}$ be the expected value of the average of \textit{n} observations of RV. Then the first moment, or “true mean”, of the random variable defined as the observed mean of \textit{n} draws on RV is $\bm{\mu}\equiv\mathbb{E}(\bm{\bar{X}})$. Therefore:

\begin{equation*}
 \mathbb{E}(\bm{V(S)}) = \mathbb{P}(\bm{\bar{X}} - \bm{\mu} \geq \bm{t})
\end{equation*}

Since $  \mathbb{E}(\bm{V(S)}) \leq \mathbb{E}(\bm{e}^{\bm{h}(\bm{S} - \mathbb{E}(\bm{S}) -\bm{nt})})$:

\begin{equation*}
 \mathbb{P}(\bm{\bar{X}} - \bm{\mu} \geq \bm{t}) \leq \mathbb{E}(\bm{e}^{\bm{h}(\bm{S} - \bm{\mathbb{E}}(\bm{S}) -\bm{nt})})
 \tag*{\textbf{A-1}}
\end{equation*}

\newpage

Since $S=\sum_{i=n}^{n}X_i$ it follows by substitution that:

\begin{equation*}
\boldsymbol{e}^{\boldsymbol{h}(S - \mathbb{E}(S))}
= \boldsymbol{e}^{\boldsymbol{h}(X_1 - \mathbb{E}(X_1))}
  \boldsymbol{e}^{\boldsymbol{h}(X_2 - \mathbb{E}(X_2))}
  \cdots
  \boldsymbol{e}^{\boldsymbol{h}(X_n - \mathbb{E}(X_n))}
= \prod_{i=1}^{n} \boldsymbol{e}^{\boldsymbol{h}(X_i - \mathbb{E}(X_i))}
\end{equation*}

Therefore: 
\begin{equation*}
\bm{e}^{\bm{h}(S - \mathbb{E}(S))} = \prod_{i=1}^{n}\bm{e}^{\bm{h}(X_i - \mathbb{E}(X_i))}
\end{equation*}

By \textbf{Assumption 2}, each observation is independent; therefore by definition, the expected value of the product of successive observations of the random variable equals the product of the expected values for each draw; in other words, for two successive draws  $\bm{X_1}$,$\bm{X_2}$, $\mathbb{E}(\bm{X_1} \cdot \bm{X_2}) = \mathbb{E}(\bm{X_2}) \cdot \mathbb{E}(\bm{X_2})$. This property holds even if the expectation value of each draw changes from draw to draw, so:

\begin{equation*}
\mathbb{E}(\bm{e}^{\bm{h}(S - \mathbb{E}(S))}) = \mathbb{E}(\prod_{i=1}^{n}\bm{e}^{\bm{h}(X_i - E(X_i))})
\end{equation*}

Multiply both sides by $e^{-hnt}$. This quantity is a constant, not a random variable; therefore $\mathbb{E}(e^{h(S - \mathbb{E}(S))}) \cdot e^{-hnt} = \mathbb{E}(e^{h(S - \mathbb{E}(S))} \cdot e^{-hnt} )$ and it follows that:

\begin{equation*}
  \bm{\mathbb{E}(e^{h(S - \mathbb{E}(S) -nt)}) = e^{-hnt}\mathbb{E}(\prod_{i=1}^{n}e^{h(X_i - \mathbb{E}(X_i))})}
 \tag*{\textbf{A-2}}
\end{equation*}

Combining \textbf{A-1} and \textbf{A-2} gives: 

\begin{equation*}
  \mathbb{P}(\bar{X} - \mu \geq t) \leq e^{-hnt}\mathbb{E}(\prod_{i=1}^{n}e^{h(X_i - \mathbb{E}(X_i))})
\end{equation*}

By the property of independence of successive draws of a random variable: 

\begin{equation*}
\mathbb{E}(\prod_{i=1}^{n}e^{h(X_i - \mathbb{E}(X_i))}) = \prod_{i=1}^{n} \mathbb{E}(e^{h X_i}e^{-h\mathbb{E}(X_i)})
\end{equation*}

By the product rule of exponentiation:

\begin{equation*}
\prod_{i=1}^{n} \mathbb{E}(e^{h X_i}e^{-h\mathbb{E}(X_i)}) = \prod_{i=1}^{n}e^{-h\mathbb{E}(X_i)}\prod_{i=1}^{n} \mathbb{E}(e^{h X_i})
\end{equation*}

$\mu_i=\mathbb{E}(X_i)$ by definition, so:

\begin{equation*}
  \prod_{i=1}^{n} e^{-h\mathbb{E}(X_i)} \prod_{i=1}^{n} \mathbb{E}(e^{hX_i})
=\prod_{i=1}^{n} e^{-h\mu_i} \prod_{i=1}^{n} \mathbb{E}(e^{hX_i})
= e^{-h(\mu_1 + \cdots + \mu_n)}\prod_{i=1}^{n} \mathbb{E}(e^ {hX_i} )
\end{equation*}

Let $n\mu = \mu_1 + \cdots + \mu_n$. Then $\mu$ may be interpreted as the “true mean of the true means of each draw of the random variable”. Representing the mean in this way is what allows the inequality to remain valid even if the mean of the random variable changes from draw to draw.  If the indices on observations of $X_i$ are time indexed, then $\mu$ represents the value expected for an observation of the RV for any unspecified time index in the range [1,\textit{n}]. 

With this definition in place: 

\begin{equation*}
e^{-h(\mu_1 + \cdots + \mu_n)} \prod_{i=1}^{n} \mathbb{E}(e^{hX_i})
= e^{-h\mu n} \prod_{i=1}^{n} \mathbb{E}(e^{hX_i}).
\end{equation*}

By the preceding chain of substitution, $\mathbb{E}\!\left(\prod_{i=1}^{n} e^{h(X_i - \mathbb{E}(X_i))}\right)= e^{-h\mu n} \prod_{i=1}^{n} \mathbb{E}(e^{hX_i})$. By substituting that equivalency into \textbf{A-2} we arrive at:

\begin{equation*}
\mathbb{P}(\bm{\bar{X} - \mu \ge t)
\le e^{-hnt} e^{-h\mu n} \prod_{i=1}^{n} \mathbb{E}(e^{hX_i})}
 \tag*{\textbf{A-3}}
\end{equation*}

To continue the proof, we now require a lemma that places upper boundaries on the exponential function $e^{hX_i}$ as follows.

\textbf{Lemma 1} and \textbf{1-2}: 

Let $X_i$ be any observation of a random variable $RV_i$ bounded by upper and lower bounds ${b}$ and ${a}$ such that $a \leq X_i \leq \bm{b}$. Let $h$ be real-valued and positive.

The equation of the straight line that connects the two points $(a, e^{ha})$ and $(b, e^{hb})$ is:

\begin{equation}
y_{\text{line}} = \frac{b - x}{b - a} e^{ha} + \frac{x - a}{b - a} e^{hb}
\end{equation}

The function $y_{\text{exp}} = e^{hx}$ is convex. By the definition of convexity, 
$y_{\text{line}}(x) \geq y_{\text{exp}}(x)$ for any $x$ that lies within the domain 
$[a, b]$. Any observation of the random variable $X_i$ lies within the domain 
$[a, b]$ because the random variable is bounded by $a$ and $b$. Therefore:

\begin{equation*}
e^{hX_i} \leq \frac{b - X_i}{b - a} e^{ha} + \frac{X_i - a}{b - a} e^{hb}
\end{equation*}

Since $h$, $a$, and $b$ are constants:

\begin{equation*}
\mathbb{E}(e^{hX_i}) \leq \frac{b - \mathbb{E}(X_i)}{b - a} e^{ha} + \frac{\mathbb{E}(X_i) - a}{b - a} e^{hb}
\end{equation*}

Since $\mu_i\equiv\mathbb{E}(X_i)$:

\begin{equation*}
\mathbb{E}\bm{(e^{hX_i})} \leq \bm{\frac{b - \mu_i}{b - a} e^{ha} + \frac{\mu_i - a}{b - a} e^{hb}}
 \tag*{\textbf{LEMMA 1}}
\end{equation*}

\textbf{Lemma 1} is presented above in a form very close to the way in which it appeared in Hoeffding’s original proof of his famous Inequality Theorems in 1963. For the current discussion, we develop 

\newpage
\textbf{Lemma 1-2} from an algebraic rearrangement of \textbf{Lemma 1}:

\begin{equation*}
\mathbb{E}(e^{hX_i}) \leq \frac{b}{b - a} e^{ha} - \frac{\mu_i}{b-a} e^{ha} + \frac{\mu_i - a}{b -a}e^{hb}
\end{equation*}

Now, is a logical fact that $b-a=b-a$ for any $a$, $b$.  It follows that $b-a+a=b$; further, that $1+\frac{a}{b-a}=\frac{b}{b-a}$; therefore, we make the substitution for $\frac{b}{b-a}$:

\begin{equation*}
\mathbb{E}(e^{hX_i}) \leq e^{ha}+\frac{a}{b-a}e^{ha}- \frac{\mu_i}{b -a}e^{ha}+ \frac{\mu_i - a}{b - a} e^{hb}
\end{equation*}

Combining terms yields \textbf{Lemma 1-2}:

\begin{equation*}
\bm{\mathbb{E}(e^{hX_i})} \leq \bm{e^{ha}} - \frac{\bm{\mu_i - a}}{\bm{b - a}} \bm{e^{ha}} + \frac{\bm{\mu_i - a}}{\bm{b - a}} \bm{e^{hb}}
 \tag*{\textbf{LEMMA 2}}
\end{equation*}

\textbf{Now we return to the proof of Hoeffding’s Inequalities.}

Making use of \textbf{Lemma 1-2}, we may substitute for $\mathbb{E}(e^{hX_i})$ in the upper bound placed on $\mathbb{P}(\bar{X}-\mu\geq t)$ in Statement \textbf{A-3} as follows:

\begin{equation*}
\mathbb{P}(\bar{X}-\mu\geq t) \leq e^{-hnt} e^{-h\mu n} \prod_{i=1}^{n} e^{ha} - \frac{\mu_i - a}{b - a} e^{ha} + \frac{\mu_i - a}{b - a} e^{hb} 
\end{equation*}

The statement above holds for all real, positive values of $h$. But since $b$ and $a$ are bounds, $b>a$ by definition, so $b-a>0$ in all cases. Therefore, the following also holds:

\begin{equation*}
\mathbb{P}(\bar{X}-\mu\geq t) \leq e^{-hn\frac{t}{b - a} } e^{-hn \frac{\mu}{b - a}} \prod_{i=1}^{n} e^{h \frac{a}{b - a}} - \frac{\mu_i - a}{b - a} e^{h \frac{a}{b - a}} + \frac{\mu_i - a}{b - a} e^{h \frac{b}{b - a}} 
\end{equation*}

Take the $n^{th}$ root of both sides:
\begin{equation*}
\sqrt[n]{\mathbb{P}(\bar{X} - \mu\geq t)} \leq e^{-h \frac{t}{b - a}} e^{-h\frac{\mu}{b - a}}\sqrt[n]{\prod_{i=1}^{n} e^{h \frac{a}{b - a}}- \frac{\mu_i - a}{b - a} e^{h \frac{a}{b - a}} + \frac{\mu_i - a}{b - a} e^{h \frac{b}{b - a}}}
\end{equation*}

The expression under the radical on the right hand side is a geometric mean. Because the geometric mean of a series is always less than or equal to the series’ arithmetic mean, we may replace the geometric mean with an arithmetic mean in the inequality as follows:

\begin{equation*}
\sqrt[n]{\mathbb{P}(\bar{X} - \mu \ge t)}\leq e^{-h \frac{t}{b - a}} e^{-h \frac{\mu}{b - a}}\frac{1}{n} \sum_{i=1}^{n}e^{h \frac{a}{b- a}} - \frac{\mu_i - a}{b - a} e^{h \frac{a}{b - a}} + \frac{\mu_i - a}{b - a} e^{h \frac{b}{b - a}} 
\end{equation*}

By the algebraic identity $\frac{b}{b-a}=1+\frac{a}{b-a}$:
\begin{equation*}
\sqrt[n]{\mathbb{P}(\bar{X} - \mu \geq t)}\leq e^{-h \frac{t}{b - a}} e^{-h \frac{\mu}{b - a}}\frac{1}{n} \sum_{i=1}^{n}e^{h \frac{a}{b - a}}-\frac{\mu_i - a}{b - a} e^{h \frac{a}{b - a}}+ \frac{\mu_i - a}{b - a} e^{h + h \frac{a}{b - a}}
\end{equation*}

And by again invoking our definition $n\mu=\mu_1+\cdots+\mu_n$ we have:
\begin{equation*}
\sqrt[n]{\mathbb{P}(\bar{X} - \mu \geq t)}\leq e^{-h \frac{t}{b - a}-\frac{\mu}{b - a}}
( e^{h \frac{a}{b - a}}- \frac{\mu - a}{b - a} e^{h \frac{a}{b - a}} + \frac{\mu - a}{b - a} e^{h + h \frac{a}{b - a}})
\end{equation*}

Raising both sides to the power of $n$ and rearranging:
\begin{equation*}
\mathbb{P}(\bar{X} - \mu \ge t)
\leq \left(
e^{-h \frac{t}{b - a}{-h \frac{\bm{\mu} - a}{b - a}}} \left( 1 - \frac{\bm{\mu} - a}{b - a} + \frac{\bm{\mu} - a}{b - a} e^{h} \right)
\right)^{n}
\end{equation*}

\textbf{Normalizing the Bounds of $\bm{\mu}$ and $\bm{t}$}

For ease of use, it is convenient to express $\mu$ and $t$ with bounds normalized to 0 and 1. We make the substitution at this point by introducing the following notation:

\begin{equation*}
\dot{\mu}\equiv \frac{\mu - a}{b - a} \quad\dot{t}\equiv \frac{t}{b - a}
\end{equation*}

The ‘dot’ on the parameters above denotes the normalization that has been applied. This notation is intended to make it clear that Hoeffding’s Inequality applies to random variables that may have \textbf{any} bounds, while keeping the simplicity of working with equations that do not require a and b appear explicitly.

\textbf{Finding \textit{h} that minimizes bounds}

With this notation in place, we arrive at Statement \textbf{A-4:}

\begin{equation*}
\mathbb{P}\bm{(\bar{X} - \mu \ge t)} \leq \left(\bm{e^{-h\dot{t}-h\dot{\mu}} (1-\dot{\mu} e^h)}
\right)^{\bm{n}}
 \tag*{\textbf{A-4}}
\end{equation*}

Critical point analysis shows that the right-hand side of \textbf{A-4 }has a minimum at

\begin{equation*}
\bm{h_0} = \text{ln} \frac{(1-\dot{\bm{\mu}})(\dot{\bm{\mu}}+\dot{t})}{(1-\dot{\bm{\mu}}-\dot{t})\dot{\bm{\mu}}}
\end{equation*}

By definition $0 \leq \dot{t} < 1 - \bm{\dot{\mu}}$. Since $\lim_{\dot{t} \to 0^{+}} \bm{h_{0}} = 0$ 
and $\lim_{\dot{t} \to (1 - \bm{\dot{\mu}})^{-}} \bm{h_{0}} = \ln \frac{1 - \bm{\dot{\mu}}}{0} = \infty$, 
$0 \leq \bm{h_{0}} < \infty$.

Substituting $\bm{h_0}$ into \textbf{A-4 }and algebraic rearrangement gives:

\begin{equation*}
\bm{\mathbb{P}(\bar{X} - \mu \ge t) \leq \left(\left(\frac{\dot{\mu}}{\dot{\mu}+\dot{t}}\right)^{\dot{\mu}+\dot{t}}\left(\frac{1-\dot{\mu}}{1-\dot{\mu}-\dot{t}}\right)^{1-\dot{\mu}-\dot{t}}\right)^{n}}
 \tag*{\textbf{Hoeffding’s Inequality }}
\end{equation*}

Note that if $\dot{t}>1-\dot{\mu}$, then the upper probability bound falls to zero. This is not simply an artifact of the construction of the statement; it’s a certainty because in order for that event to take place, the random variable would have to exceed the bounds \textit{a} and \textit{b}.

\newpage 
\textbf{Hoeffding’s Inequality, Exponential form:}

The mathematical statement usually referred to as Hoeffding’s Inequality, and referenced earlier in this paper, has the following form: 

{\boldmath
\begin{equation}
\label{eq:hoeffding2}
\mathbb{P}[E_{in} - E_{out} \geq t] \leq e^{-2t^{2}n}
\tag*{\textbf{Hoeffding’s Inequality 2}}
\end{equation}
\unboldmath
}

This statement was intended as a simplification of \textbf{Hoeffding’s Inequality 1} as derived in the previous section. It holds because the bounds given by the right-hand side of the inequality always exceed the bounds given by the right-hand side of \textbf{Inequality 1} for any combination of $\mu$,$t$ within the bounds $a$ and $b$, and does not require the modeler to specify $\mu$ to calculate the bound on the right-hand side.

\clearpage 
\begingroup      
\normalsize 
\setlength{\parindent}{0pt}
\setlength{\parskip}{0pt}

\urlstyle{same}
\renewcommand{\UrlFont}{\normalfont}
\nocite{AbuMostafa2012}
\printbibliography

\end{document}